\documentclass[prl,aps,superscriptaddress,amsmath,amssymb,showpacs,twocolumn,floatfix]{revtex4}
\usepackage[pdftitle={Charge density wave surface phase slips and non-contact nanofriction}]{hyperref}
\usepackage{epsfig}
\usepackage{amsfonts}
\usepackage{amsmath}
\usepackage{natbib}
\newcommand{\md}[1]{\left|#1\right|}
\newcommand{\FE}{\mathcal{F}}
\newcommand{\qq}{\mathbf{Q}}
\newcommand{\rr}{\mathbf{r}}

\begin{document}

\title{Charge density wave surface phase slips and non-contact nanofriction}

\author{Franco Pellegrini}
\affiliation{SISSA, Via Bonomea 265, I-34136 Trieste, Italy}
\affiliation{CNR-IOM Democritos National Simulation Center, Via Bonomea 265, I-34136 Trieste, Italy}

\author{Giuseppe E. Santoro}
\affiliation{SISSA, Via Bonomea 265, I-34136 Trieste, Italy}
\affiliation{CNR-IOM Democritos National Simulation Center, Via Bonomea 265, I-34136 Trieste, Italy}
\affiliation{International Centre for Theoretical Physics (ICTP), P.O. Box 586, I-34014 Trieste, Italy}

\author{Erio Tosatti}
\affiliation{SISSA, Via Bonomea 265, I-34136 Trieste, Italy}
\affiliation{CNR-IOM Democritos National Simulation Center, Via Bonomea 265, I-34136 Trieste, Italy}
\affiliation{International Centre for Theoretical Physics (ICTP), P.O. Box 586, I-34014 Trieste, Italy}

\date{\today}

\begin{abstract}
Bulk electrical dissipation caused by charge-density-wave (CDW)  depinning and sliding is a classic subject. 
We present a novel local, nanoscale mechanism describing the occurrence of mechanical dissipation 
peaks in the dynamics of an atomic force microscope tip oscillating above the surface of a CDW material. 
Local surface 2$\pi$ slips of the CDW phase are predicted to take place giving rise to mechanical hysteresis
and large dissipation at discrete tip surface distances. 
The results of our static and dynamic numerical simulations are believed to be relevant to recent experiments on NbSe$_2$;
other candidate systems in which similar effects should be observable are also discussed.
\end{abstract}


\maketitle


Charge-density-waves (CDWs) are static modulations of small amplitude and generally incommensurate
periodicities which occur in the electron density distribution and in the lattice positions of a variety of  
materials~\cite{Gruner_RMP88}. They may derive either by an exchange-driven instability of a metallic 
Fermi surface~\cite{Overhauser_PR68},  or by a lattice dynamical 
instability leading to a static periodic lattice distortion (PLD) which may equivalently be driven by electrons 
near Fermi~\cite{Peierls_55,Woll_PR62} or just by anharmonicity~\cite{Weber_PRL11}.  
A CDW superstructure, characterized by amplitude $\rho_0$ and phase $\phi(x)$ relative to the underlying 
crystal lattice can be made to slide with transport of mass and charge and with energy dissipation under 
external perturbations and fields~\cite{Gruner_RMP88}. 

Phase slips in bulk CDWs/PLDs are involved in a variety of phenomena, including noise 
generation~\cite{Coppersmith_PRL90}, switching~\cite{Inui_PRB88}, current conversion at contacts~\cite{Maher_PRL92}, 
noise~\cite{Ong_PRL84,Gruner_PRL81} and more. While these phenomena are now classic knowledge, 
there is to date no parallel work addressing the possibility to mechanically provoke CDW phase slips 
at a chosen local point. In this letter we describe a two-dimensional model showing how a localized CDW/PLD 
phase slip may be provoked by external action of an atomic force microscope (AFM) tip at an arbitrarily 
chosen point outside a surface.  

Experiments have for some time revealed the dissipative and frictional effects experienced by nanoprobes in 
contact or near contact with different surfaces, and considerable theoretical effort is being devoted to their 
understanding~\cite{Vanossi_RMP13}. The development of ultra-sensitive tools such as  the ``pendulum'' 
AFM~\cite{Stipe_PRL01,Gysin_RSI11} offers a chance 
to investigate more delicate and intimate substrate properties. Near a CDW material the tip oscillations may 
actuate, through van der Waals or electrostatic coupling,  an electronic and atomic movement in the surface 
right under the tip, amounting in this case to coupling to the CDW order parameter. 
Owing to the periodic nature of the CDW state, the coupled tip-CDW system has multiple solutions, characterized 
by a different winding number (a topological property) which differ by a local phase slip, and correspond 
to different energy branches. At the precise tip-surface distance where the two branches cross, 
the system will jump from one to the other injecting a local 2$\pi$ phase slip, and the corresponding hysteresis 
cycle will reflect directly as a mechanical dissipation, persisting even at low tip oscillation frequencies. 
This scenario and these results are believed to represent closely what is going on in recent experiments on 
the CDW material NbSe$_2$~\cite{Langer_NatMat14}.
 
\textit{The Model} --- Irrespective of the microscopic mechanism that generated it, we introduce the CDW as a 
periodic modulation of the ion and electron density $\rho$, of the form  $\Delta\rho(\rr)=\rho_0\cos(\qq\cdot\rr+\phi_0)$, 
where $\rho_0$ is the amplitude, $\lambda \sim 2\pi Q^{-1}$ the characteristic wavelength, and $\phi_0$ an initially
constant phase, fixed by some far away agent. We wish to study the effect of a localized perturbation represented by
a weakly interacting and slowly oscillating nano or mesoscopic sized probe hovering above the surface, acting on a 
length scale $\sigma$ similar to the CDW wavelength, $\sigma\sim \lambda$.
In the past, uniform perturbations such as external electric fields or point-like perturbations such as 
pinning defects have been studied~\cite{Gruner_RMP88,Fukuyama_PRB78,Lee_PRB79,Tucker_PRB89}, describing global
CDW dynamical sliding, or local static CDW pinning. To address the problem of local CDW dynamics, we now go beyond
the straight one-dimensional approximations commonly used in the past.

As in the standard Fukuyama-Lee-Rice model~\cite{Fukuyama_PRB78,Lee_PRB79} we treat the CDW 
at the Ginzburg-Landau level as a classical elastic medium, where the CDW modulation is described by a static 
space (and, later, time) dependent order parameter
$\Delta\rho(\rr)=A(\rr)\cos(\qq\cdot\rr+\phi(\rr))$. The unperturbed CDW has constant $A(\rr)=\rho_0$ 
and $\phi(\rr)=\phi_0$ and the free energy reads:
\begin{equation} \label{eq:FE0}
\FE_0[\psi(\rr)]=\int\left[-2f_0\md{\psi(\rr)}^2+f_0\md{\psi(\rr)}^4+\kappa\md{\nabla\psi(\rr)}^2\right]\mathrm{d}\rr
\;,
\end{equation}
where $\psi(\rr)=A(\rr) e^{i\phi(\rr)}$, $f_0$ and $\kappa$ are nonlinearity and stiffness real positive parameters, 
respectively, and a unidirectional CDW modulation has been assumed (three superposed modulations could equally well be treated).   
Next, the perturbation induced by an AFM tip is described as a potential $V(\rr)$ coupling to the order parameter
\begin{equation} \label{eq:FEV}
\FE_V[\psi(\rr)]=\int V(\rr) \mathrm{Re} \left[\psi(\rr)e^{i\qq\cdot\rr} \right] \mathrm{d}\rr \;.
\end{equation}
Past studies of point impurities~\cite{Tucker_PRB89,Tutto_PRB85} assumed $V(\rr)=\sum_i\delta(\rr-\rr_i)$, one dimension, 
and coupling was restricted to the CDW phase only, but all of that is inadequate here. 
Indeed, if one considers a phase-only functional of the form
\begin{equation} \label{eq:FEphi}
\FE_{\phi}[\phi(\rr)] = 
\int\left[\kappa\md{\nabla\phi(\rr)}^2+V(\rr)\rho_0\cos(\qq\cdot\rr+\phi(\rr))\right]\mathrm{d}\rr
\end{equation} 
and minimizes it in one dimension, the result will be a linear behavior of $\phi(x)$ away from the impurity, unphysical.
Therefore, it is mandatory to work in at least two dimensions, where the Laplacian can accommodate
solutions which decay far from the perturbation.
Moreover, the nature of the phase, defined modulo $2\pi$, implies that, given some boundary conditions, 
the solution is not uniquely defined unless we also specify the total variation of $\phi$. 
Assuming the phase to have the unperturbed value $\phi_0$ far from the perturbation, we can define 
the \textit{winding-number} $N$ 
\begin{equation} \label{eq:WN}
N=\frac{1}{2\pi}\int\! \nabla\phi(\rr)\; \mathrm{d}x \;,
\end{equation}
taken along the CDW direction $\qq$. The winding-number $N$ is needed to fully specify a solution.
However, this procedure leads to a problem of the phase-only approximation \eqref{eq:FEphi}: any change in the winding 
number along the $\qq$ direction will reflect itself in the whole perpendicular direction, thus unphysically
raising the energy of such a solution. For a physically sensible result we need to involve the amplitude 
degree of freedom, which will allow for the presence of dislocations and for local changes in the winding-number.
We will therefore consider $V(\rr)$ with a finite width $\sigma$ of the order of the wavelength $\lambda$,
and minimize the total phase and amplitude dependent free energy $\FE=\FE_0+\FE_V$ given a specific shape of $V(\rr)$. 
The final result is expected to be similar to what previously considered in the wider context of 
phase-slip~\cite{Tucker_PRB89} and more specifically in the case of localized phase-slip 
centers~\cite{Maki_PLA95,Gorkov_ZETF84}. 
Namely, the local strain induced by the perturbation on the phase will reduce the order parameter amplitude,
to the point where a local phase-slip event becomes possible. In more than one dimension, the boundary 
between areas with different winding-number will be marked by structures such as \textit{vortices}.

\begin{figure}[!tb]\centering
\includegraphics[width=.5\textwidth]{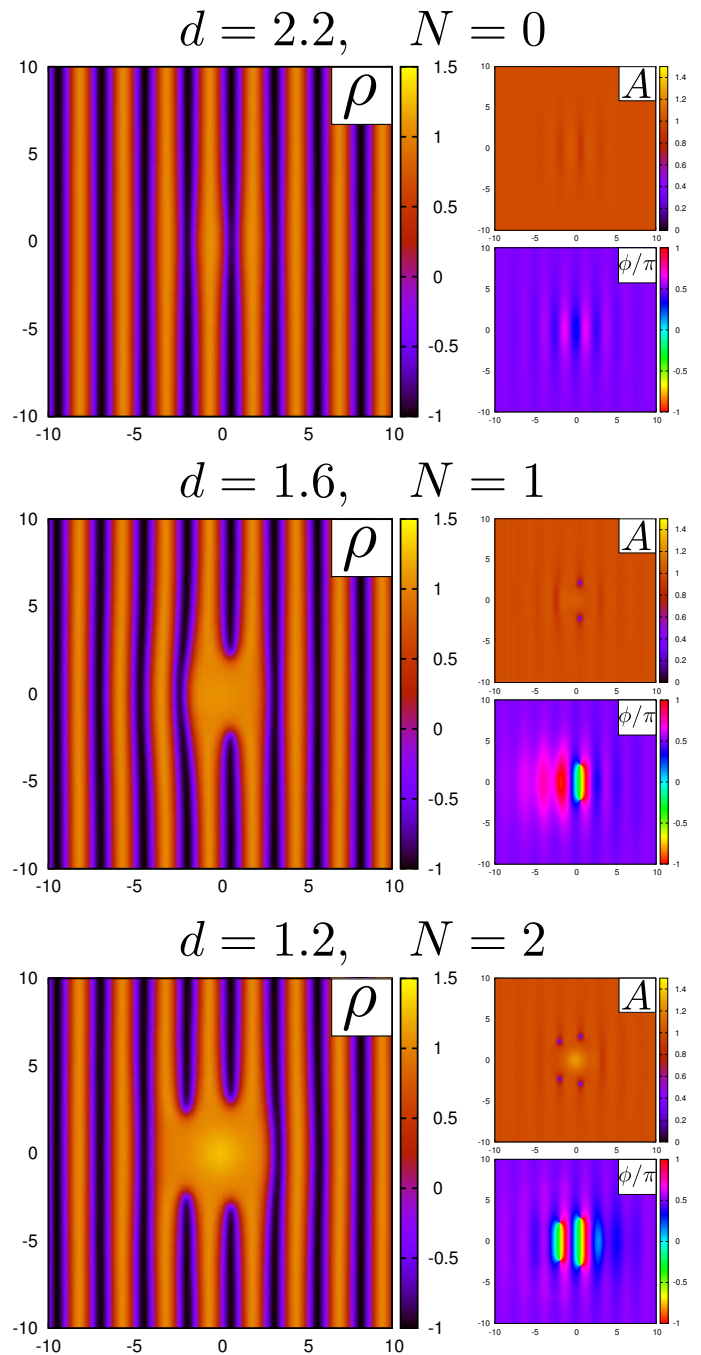}
\caption{\label{fig:minima} Charge density $\rho$, order parameter amplitude $A$ and phase $\phi$  potraits for 
minimal free energy solutions with different winding-number $N$ and tip-surface distance $d$ (in nm). Results 
from simulations on a $201\times201$ grid with parameters (see text) $f_0=2$ eV/nm, $\kappa=0.2$ eV, $Q=2.5$ nm, 
$\bar{V}=-9.4$ eV$\cdot$nm, $\bar{\sigma}=1,2$ nm$^{-1}$ and boundary conditions $\psi_0=i$ (right and left 
sides).}
\end{figure}

From this preliminary analysis we can now anticipate that, as the tip approaches the surface, increasing
the strength of $V(\rr)$, it might reach points where the energies of solutions with different winding-numbers 
cross. At these points the transition between successive winding-number manifolds would not be continuous, 
due to the barrier required to create the vortices. As a result, time dependent oscillations of the tip around 
these locations would generally occur with hysteresis and, ultimately,  mechanical dissipation despite  low 
oscillation frequencies.

\textit{Simulations} --- 
To verify the proposed mechanism, we performed numerical simulations of the tip-CDW surface model, 
restricted to two dimensions assuming that all effects will heal out below the surface (reasonable in a layer compound).
To mimic the tip potential we integrate a van der Waals potential $C/r^6$ over a conical shape at distance $d$ 
from the surface. The result can be reasonably approximated in the main area under the tip as a Lorentzian 
\begin{equation} \label{eq:ExtPot}
V_d(\rr)=\frac{V_d^{0}}{\rr^2+\sigma_d^2} \;,
\end{equation}
where $\rr$ is the in-plane distance from the tip central axis, 
and the parameters scale like $V_d^{0}=\bar{V}/d$ and $\sigma_d=\bar{\sigma}d^2$.

\begin{figure}[!tb]\centering
\includegraphics[width=.45\textwidth]{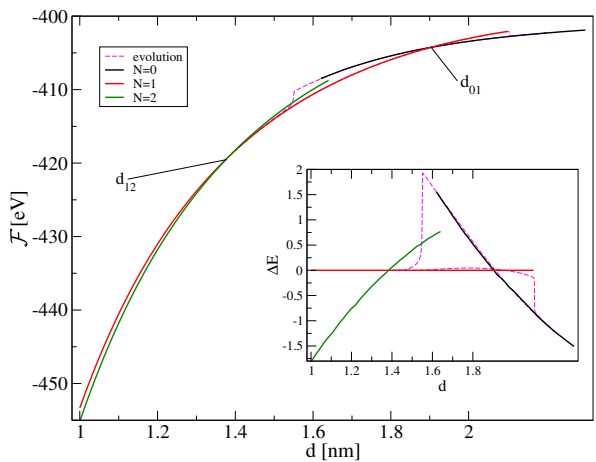}
\caption{\label{fig:encurves}
Minimal free energy $\FE$ as a function of tip distance $d$ for subspaces with different winding number $N$ 
(full lines) and evolution of $\FE$ during an oscillation with $d_0=1.8$ nm, $\bar{d}=0.4$ nm, 
$\omega=6\cdot 10^4$ Hz (other parameters are the same as Fig.~\ref{fig:minima}). Inset: same data rescaled 
with respect to the $N=1$ value to highlight their difference.} 
\end{figure}

To minimize the total free energy $\FE=\FE_0+\FE_V$, we discretize the complex order parameter $\psi$ on a 
square grid of points with a spacing much smaller than 
the characteristic wavelength of the CDW, and impose a constant boundary condition $\psi_0$ on the sides 
perpendicular to $\qq$, while setting periodic boundary conditions in the direction parallel to $\qq$ to allow for 
possible phase jumps~\cite{CDW_note1}.  
The minimization is carried out with a standard conjugated gradients algorithm~\cite{NumRec_07}.
Without aiming at a numerically realistic representation of any experimental system, we use reasonable order 
of magnitude estimates of the system parameters, that help us build a clear if qualitative portrait of the 
tip-induced CDW phase slip.

Fig.~\ref{fig:minima} shows the order parameter modulation 
amplitude $\rho$ and phase $\phi$ for 
minima with different winding-number $N$, for a non-contact (attractive) tip at different distances $d$. 
The winding-number is calculated along the line passing through the point right below the tip 
(center of the simulation cell) according to Eq.~\eqref{eq:WN}, with $N=0$ being the unperturbed case. 
As predicted, we see upon decreasing $d$ through the first and successive critical distances $d_{01}, d_{12}$, etc.  
the appearance of a pair of vortices (with opposite rotation) for every unit increase of the winding number. 
These vortices are characterized by a suppression of the amplitude and a total change of the phase by $2\pi$ 
on a path around them, since they separate the phase-slippage center from the unaffected area far 
from the tip.

Since the solution with a given winding number lies in a local minimum, it is possible to use the minimization 
algorithm, for example  by starting from a reasonable configuration, 
to find solutions in a certain subspace, even when that is not the global minimum for that given case. 
This allows us to extend the calculation of the local free energy minima in a given $N$ subspace well beyond 
their crossing points, generating a family of free energy curves of definite $N$ as a function of the distance $d$. 
Fig.~\ref{fig:encurves} (full lines) is an example, showing two successive crossing points. We expect each crossing 
to give rise to a first order transition, and thus to a hysteretic peak in the experimental dissipation trace. Of course, a 
more complex CDW configuration or different parameters could give rise to more and different peaks.

\begin{figure}[!b]\centering
\includegraphics[width=.45\textwidth]{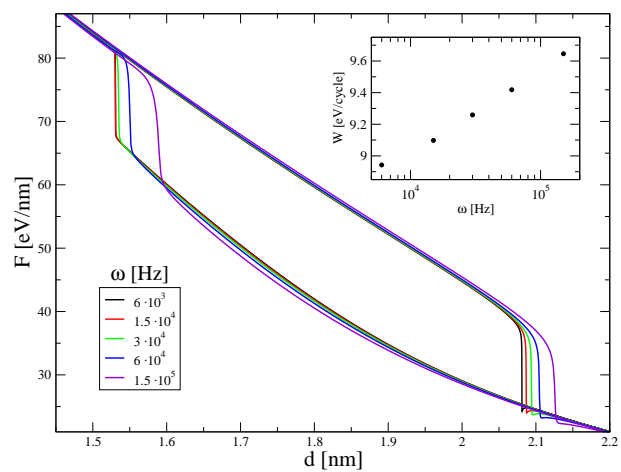}
\caption{\label{fig:fcurves}
Force as a function of distance for evolutions with $d_0=1.8$ nm, $\bar{d}=0.4$ nm and different values of
$\omega$ with a coefficient $\Gamma=10^{-7}$ eV$\cdot$s. Inset: total work $W$ as a function of oscillation 
frequency $\omega$.}
\end{figure}

In order to demonstrate the phenomenon in full, we now extend our study to the tip-CDW
dynamics. To this end, we carry out a simulated evolution generated by the time-dependent 
Ginzburg-Landau equation~\cite{Gorkov_ZETF84}
\begin{equation} \label{eq:TDGL}
-\Gamma\frac{\partial\psi}{\partial t}=\frac{\delta\FE}{\delta\psi^*} \,
\end{equation}
which can be interpreted as an overdamped evolution of the order parameter towards the 
equilibrium position, with a damping coefficient $\Gamma$. 
Integrating this equation (through a standard Runge-Kutta algorithm~\cite{NumRec_07}) we compute the 
time evolution of the free energy, as shown by the dotted line in Fig.~\ref{fig:encurves}, for a tip performing 
a full oscillation perpendicular to the surface of the form $d(t)=d_0+\bar{d}\cos(\omega t)$.
The result shows that in this effectively adiabatic evolution the system is stuck in its winding-number manifold 
well beyond the crossing point, effectively realizing a hysteresis cycle. 
Fig.~\ref{fig:fcurves} shows the force hysteresis for oscillations at different frequencies $\omega$: 
the area of the cycles directly represents the dissipated energy per cycle $W$,  given in the inset.
It should be stressed here that what we calculated is in effect only a {\it maximal} hysteresis cycle. The actual size of the 
cycle, and thus of the total dissipated tip mechanical energy, is in principle smaller (and should in effect vanish in the limit 
of vanishing oscillation frequency) due to thermal fluctuations which our treatment omits. However, as known
in other cases, the large, mesoscopic size of the tip-surface mechanics generally makes the simple adiabatic description rather 
accurate, and the effective hysteresis only modestly frequency and temperature dependent. 

\textit{Discussion and conclusions} --- We have shown that local surface CDW phase slips and vortex pairs can be introduced by the external
potential of an approaching tip.  In the context of macroscopic CDW conduction noise~\cite{Ong_PRL84,Gruner_PRL81,Gruner_RMP88}, 
the creation and movement of vortices has been invoked earlier in connection with phase slips near the CDW boundaries. 
In a broader context, our system can be placed in between these macroscopic situations and the simple  
models of defect pinning and phase-slip~\cite{Tucker_PRB89} by a localized perturbation.

Experimentally, Langer et al.~\cite{Langer_NatMat14} recently reported AFM dissipation peaks appearing at discrete
tip-surface distances above the CDW material 2H-NbSe$_2$, qualitatively suggesting in a 1D model 
the injection of $2\pi$ phase slips. The present results describe at the minimal level a theory 
that can explain this type of phenomenon, connecting the phase slip to a vortex pair formation, and
providing the time dependent portrait of the injection process.

It would be of considerable interest in the future to explore further this effect in other systems with different characteristics.
In insulating, quasi-one dimensional CDW systems the injected phase slip should also amount to
the injection of a quantized, possibly fractional pairs of opposite charges~\cite{CDW_note2}. 
In a spin density wave system, such as the chromium surface,  a nonmagnetic tip would still couple 
to the accompanying CDW~\cite{Kim_PRB13} where surface phase slips could be injected.
In superconductors, the induction of single vortices over Pb thin film islands has been experimentally
verified \cite{Hasegawa_Nanotec10} and the feasibility of controlling single vortices through magnetic force
microscopy (MFM) tips demonstrated \cite{Auslander_NatPhy09}: it would be interesting to probe for
dissipation peaks, as we have addressed above, induced by the MFM tip creation of vortex pairs in thin superconducting films.

\textit{Acknowledgements} --- 
This project was carried out under the ERC Advanced Research Grant N. 320796 MODPHYSFRICT. We are especially
indebted to our colleagues of SINERGIA SNSF Project CRSII2 136287/1 for support, and for exchange of information and ideas.  
We also acknowledge general research support by MIUR, through PRIN-2010LLKJBX\_001. 


\begin{thebibliography}{26}
\expandafter\ifx\csname natexlab\endcsname\relax\def\natexlab#1{#1}\fi
\expandafter\ifx\csname bibnamefont\endcsname\relax
  \def\bibnamefont#1{#1}\fi
\expandafter\ifx\csname bibfnamefont\endcsname\relax
  \def\bibfnamefont#1{#1}\fi
\expandafter\ifx\csname citenamefont\endcsname\relax
  \def\citenamefont#1{#1}\fi
\expandafter\ifx\csname url\endcsname\relax
  \def\url#1{\texttt{#1}}\fi
\expandafter\ifx\csname urlprefix\endcsname\relax\def\urlprefix{URL }\fi
\providecommand{\bibinfo}[2]{#2}
\providecommand{\eprint}[2][]{\url{#2}}

\bibitem[{\citenamefont{Gr{\"u}ner}(1988)}]{Gruner_RMP88}
\bibinfo{author}{\bibfnamefont{G.}~\bibnamefont{Gr{\"u}ner}},
  \bibinfo{journal}{Rev. Mod. Phys.} \textbf{\bibinfo{volume}{60}},
  \bibinfo{pages}{1129} (\bibinfo{year}{1988}).

\bibitem[{\citenamefont{Overhauser}(1968)}]{Overhauser_PR68}
\bibinfo{author}{\bibfnamefont{A.~W.} \bibnamefont{Overhauser}},
  \bibinfo{journal}{Phys. Rev.} \textbf{\bibinfo{volume}{167}},
  \bibinfo{pages}{691} (\bibinfo{year}{1968}).

\bibitem[{\citenamefont{Peierls}(1955)}]{Peierls_55}
\bibinfo{author}{\bibfnamefont{R.~E.} \bibnamefont{Peierls}},
  \emph{\bibinfo{title}{Quantum Theory of Solids}} (\bibinfo{publisher}{Oxford
  University Press}, \bibinfo{year}{1955}).

\bibitem[{\citenamefont{Woll and Kohn}(1962)}]{Woll_PR62}
\bibinfo{author}{\bibfnamefont{E.~J.} \bibnamefont{Woll}} \bibnamefont{and}
  \bibinfo{author}{\bibfnamefont{W.}~\bibnamefont{Kohn}},
  \bibinfo{journal}{Phys. Rev.} \textbf{\bibinfo{volume}{126}},
  \bibinfo{pages}{1693} (\bibinfo{year}{1962}).

\bibitem[{\citenamefont{Weber et~al.}(2011)\citenamefont{Weber, Rosenkranz,
  Castellan et~al.}}]{Weber_PRL11}
\bibinfo{author}{\bibfnamefont{F.}~\bibnamefont{Weber}},
  \bibinfo{author}{\bibfnamefont{S.}~\bibnamefont{Rosenkranz}},
  \bibinfo{author}{\bibfnamefont{J.-P.} \bibnamefont{Castellan}},
  \bibnamefont{\textit{et~al.}},
  \bibinfo{journal}{Phys. Rev. Lett.} \textbf{\bibinfo{volume}{107}},
  \bibinfo{pages}{107403} (\bibinfo{year}{2011}).

\bibitem[{\citenamefont{Coppersmith}(1990)}]{Coppersmith_PRL90}
\bibinfo{author}{\bibfnamefont{S.~N.} \bibnamefont{Coppersmith}},
  \bibinfo{journal}{Phys. Rev. Lett.} \textbf{\bibinfo{volume}{65}},
  \bibinfo{pages}{1044} (\bibinfo{year}{1990}).

\bibitem[{\citenamefont{Inui et~al.}(1988)\citenamefont{Inui, Hall, Doniach,
  and Zettl}}]{Inui_PRB88}
\bibinfo{author}{\bibfnamefont{M.}~\bibnamefont{Inui}},
  \bibinfo{author}{\bibfnamefont{R.~P.} \bibnamefont{Hall}},
  \bibinfo{author}{\bibfnamefont{S.}~\bibnamefont{Doniach}}, \bibnamefont{and}
  \bibinfo{author}{\bibfnamefont{A.}~\bibnamefont{Zettl}},
  \bibinfo{journal}{Phys. Rev. B} \textbf{\bibinfo{volume}{38}},
  \bibinfo{pages}{13047} (\bibinfo{year}{1988}).

\bibitem[{\citenamefont{Maher et~al.}(1992)\citenamefont{Maher, Adelman,
  Ramakrishna et~al.}}]{Maher_PRL92}
\bibinfo{author}{\bibfnamefont{M.~P.} \bibnamefont{Maher}},
  \bibinfo{author}{\bibfnamefont{T.~L.} \bibnamefont{Adelman}},
  \bibinfo{author}{\bibfnamefont{S.}~\bibnamefont{Ramakrishna}},
  \bibnamefont{\textit{et~al.}}, \bibinfo{journal}{Phys. Rev. Lett.}
  \textbf{\bibinfo{volume}{68}}, \bibinfo{pages}{3084} (\bibinfo{year}{1992}).

\bibitem[{\citenamefont{Ong et~al.}(1984)\citenamefont{Ong, Verma, and
  Maki}}]{Ong_PRL84}
\bibinfo{author}{\bibfnamefont{N.~P.} \bibnamefont{Ong}},
  \bibinfo{author}{\bibfnamefont{G.}~\bibnamefont{Verma}}, \bibnamefont{and}
  \bibinfo{author}{\bibfnamefont{K.}~\bibnamefont{Maki}},
  \bibinfo{journal}{Phys. Rev. Lett.} \textbf{\bibinfo{volume}{52}},
  \bibinfo{pages}{663} (\bibinfo{year}{1984}).

\bibitem[{\citenamefont{Gr{\"u}ner et~al.}(1981)\citenamefont{Gr{\"u}ner,
  Zawadowski, and Chaikin}}]{Gruner_PRL81}
\bibinfo{author}{\bibfnamefont{G.}~\bibnamefont{Gr{\"u}ner}},
  \bibinfo{author}{\bibfnamefont{A.}~\bibnamefont{Zawadowski}},
  \bibnamefont{and} \bibinfo{author}{\bibfnamefont{P.~M.}
  \bibnamefont{Chaikin}}, \bibinfo{journal}{Phys. Rev. Lett.}
  \textbf{\bibinfo{volume}{46}}, \bibinfo{pages}{511} (\bibinfo{year}{1981}).

\bibitem[{\citenamefont{Vanossi et~al.}(2013)\citenamefont{Vanossi, Manini,
  Urbakh, Zapperi, and Tosatti}}]{Vanossi_RMP13}
\bibinfo{author}{\bibfnamefont{A.}~\bibnamefont{Vanossi}},
  \bibinfo{author}{\bibfnamefont{N.}~\bibnamefont{Manini}},
  \bibinfo{author}{\bibfnamefont{M.}~\bibnamefont{Urbakh}},
  \bibinfo{author}{\bibfnamefont{S.}~\bibnamefont{Zapperi}}, \bibnamefont{and}
  \bibinfo{author}{\bibfnamefont{E.}~\bibnamefont{Tosatti}},
  \bibinfo{journal}{Rev. Mod. Phys.} \textbf{\bibinfo{volume}{85}},
  \bibinfo{pages}{529} (\bibinfo{year}{2013}).

\bibitem[{\citenamefont{Stipe et~al.}(2001)\citenamefont{Stipe, Mamin, Stowe,
  Kenny, and Rugar}}]{Stipe_PRL01}
\bibinfo{author}{\bibfnamefont{B.~C.} \bibnamefont{Stipe}},
  \bibinfo{author}{\bibfnamefont{H.~J.} \bibnamefont{Mamin}},
  \bibinfo{author}{\bibfnamefont{T.~D.} \bibnamefont{Stowe}},
  \bibinfo{author}{\bibfnamefont{T.~W.} \bibnamefont{Kenny}}, \bibnamefont{and}
  \bibinfo{author}{\bibfnamefont{D.}~\bibnamefont{Rugar}},
  \bibinfo{journal}{Phys. Rev. Lett.} \textbf{\bibinfo{volume}{87}},
  \bibinfo{pages}{096801} (\bibinfo{year}{2001}).

\bibitem[{\citenamefont{Gysin et~al.}(2011)\citenamefont{Gysin, Rast, Kisiel,
  Werle, and Meyer}}]{Gysin_RSI11}
\bibinfo{author}{\bibfnamefont{U.}~\bibnamefont{Gysin}},
  \bibinfo{author}{\bibfnamefont{S.}~\bibnamefont{Rast}},
  \bibinfo{author}{\bibfnamefont{M.}~\bibnamefont{Kisiel}},
  \bibinfo{author}{\bibfnamefont{C.}~\bibnamefont{Werle}}, \bibnamefont{and}
  \bibinfo{author}{\bibfnamefont{E.}~\bibnamefont{Meyer}},
  \bibinfo{journal}{Rev. Sci. Instrum.} \textbf{\bibinfo{volume}{82}},
  \bibinfo{pages}{023705} (\bibinfo{year}{2011}).

\bibitem[{\citenamefont{Langer et~al.}(2014)\citenamefont{Langer, Kisiel,
  Pawlak et~al.}}]{Langer_NatMat14}
\bibinfo{author}{\bibfnamefont{M.}~\bibnamefont{Langer}},
  \bibinfo{author}{\bibfnamefont{M.}~\bibnamefont{Kisiel}},
  \bibinfo{author}{\bibfnamefont{R.}~\bibnamefont{Pawlak}},
  \bibnamefont{\textit{et~al.}}, \bibinfo{journal}{Nature Mat.}
  \textbf{\bibinfo{volume}{13}}, \bibinfo{pages}{173} (\bibinfo{year}{2014}).

\bibitem[{\citenamefont{Fukuyama and Lee}(1978)}]{Fukuyama_PRB78}
\bibinfo{author}{\bibfnamefont{H.}~\bibnamefont{Fukuyama}} \bibnamefont{and}
  \bibinfo{author}{\bibfnamefont{P.~A.} \bibnamefont{Lee}},
  \bibinfo{journal}{Phys. Rev. B} \textbf{\bibinfo{volume}{17}},
  \bibinfo{pages}{535} (\bibinfo{year}{1978}).

\bibitem[{\citenamefont{Lee and Rice}(1979)}]{Lee_PRB79}
\bibinfo{author}{\bibfnamefont{P.~A.} \bibnamefont{Lee}} \bibnamefont{and}
  \bibinfo{author}{\bibfnamefont{T.~M.} \bibnamefont{Rice}},
  \bibinfo{journal}{Phys. Rev. B} \textbf{\bibinfo{volume}{19}},
  \bibinfo{pages}{3970} (\bibinfo{year}{1979}).

\bibitem[{\citenamefont{Tucker}(1989)}]{Tucker_PRB89}
\bibinfo{author}{\bibfnamefont{J.~R.} \bibnamefont{Tucker}},
  \bibinfo{journal}{Phys. Rev. B} \textbf{\bibinfo{volume}{40}},
  \bibinfo{pages}{5447} (\bibinfo{year}{1989}).

\bibitem[{\citenamefont{T{\"u}tt{\H{o}} and Zawadowski}(1985)}]{Tutto_PRB85}
\bibinfo{author}{\bibfnamefont{I.}~\bibnamefont{T{\"u}tt{\H{o}}}}
  \bibnamefont{and}
  \bibinfo{author}{\bibfnamefont{A.}~\bibnamefont{Zawadowski}},
  \bibinfo{journal}{Phys. Rev. B} \textbf{\bibinfo{volume}{32}},
  \bibinfo{pages}{2449} (\bibinfo{year}{1985}).

\bibitem[{\citenamefont{Maki}(1995)}]{Maki_PLA95}
\bibinfo{author}{\bibfnamefont{K.}~\bibnamefont{Maki}}, \bibinfo{journal}{Phys.
  Lett. A} \textbf{\bibinfo{volume}{202}}, \bibinfo{pages}{313}
  (\bibinfo{year}{1995}).

\bibitem[{\citenamefont{Gor'kov}(1984)}]{Gorkov_ZETF84}
\bibinfo{author}{\bibfnamefont{L.~P.} \bibnamefont{Gor'kov}},
  \bibinfo{journal}{Zh. Eksp. Teor. Fiz.} \textbf{\bibinfo{volume}{86}},
  \bibinfo{pages}{1818} (\bibinfo{year}{1984}).

\bibitem[{CDW({\natexlab{a}})}]{CDW_note1}
\bibinfo{note}{These overall boundary conditions are meant to effectively
  represent the situation of a real system such as NbSe$_2$, to be addressed
  later, exhibiting large regions of clean incommensurate CDW, whose phase is
  fixed by pinning agents such as defects at infinity}.

\bibitem[{\citenamefont{Press et~al.}(2007)\citenamefont{Press, Teukolsky,
  Vetterling, and Flannery}}]{NumRec_07}
\bibinfo{author}{\bibfnamefont{W.~H.} \bibnamefont{Press}},
  \bibinfo{author}{\bibfnamefont{S.~A.} \bibnamefont{Teukolsky}},
  \bibinfo{author}{\bibfnamefont{W.~T.} \bibnamefont{Vetterling}},
  \bibnamefont{and} \bibinfo{author}{\bibfnamefont{B.~P.}
  \bibnamefont{Flannery}}, \emph{\bibinfo{title}{Numerical Recipes: The Art of
  Scientific Computing (3rd ed.)}} (\bibinfo{publisher}{Cambridge University
  Press}, \bibinfo{year}{2007}).

\bibitem[{CDW({\natexlab{b}})}]{CDW_note2}
\bibinfo{note}{This feature is not expected in NbSe$_2$, which is metallic and
  in reality an anharmonicity-driven PLD~\cite{Weber_PRL11}. Better candidates
  for these effects would be NbSe$_3$ or TaS$_3$}.

\bibitem[{\citenamefont{Kim et~al.}(2013)\citenamefont{Kim, Logan, Shpyrko,
  Littlewood, and Isaacs}}]{Kim_PRB13}
\bibinfo{author}{\bibfnamefont{H.~C.} \bibnamefont{Kim}},
  \bibinfo{author}{\bibfnamefont{J.~M.} \bibnamefont{Logan}},
  \bibinfo{author}{\bibfnamefont{O.~G.} \bibnamefont{Shpyrko}},
  \bibinfo{author}{\bibfnamefont{P.~B.} \bibnamefont{Littlewood}},
  \bibnamefont{and} \bibinfo{author}{\bibfnamefont{E.~D.}
  \bibnamefont{Isaacs}}, \bibinfo{journal}{Phys. Rev. B}
  \textbf{\bibinfo{volume}{88}}, \bibinfo{pages}{140101(R)}
  (\bibinfo{year}{2013}).

\bibitem[{\citenamefont{Nishio et~al.}(2010)\citenamefont{Nishio, Lin, An,
  Eguchi, and Hasegawa}}]{Hasegawa_Nanotec10}
\bibinfo{author}{\bibfnamefont{T.}~\bibnamefont{Nishio}},
  \bibinfo{author}{\bibfnamefont{S.}~\bibnamefont{Lin}},
  \bibinfo{author}{\bibfnamefont{T.}~\bibnamefont{An}},
  \bibinfo{author}{\bibfnamefont{T.}~\bibnamefont{Eguchi}}, \bibnamefont{and}
  \bibinfo{author}{\bibfnamefont{Y.}~\bibnamefont{Hasegawa}},
  \bibinfo{journal}{Nanotechnology} \textbf{\bibinfo{volume}{21}},
  \bibinfo{pages}{465704} (\bibinfo{year}{2010}).

\bibitem[{\citenamefont{Auslaender et~al.}(2009)\citenamefont{Auslaender, Luan,
  Straver et~al.}}]{Auslander_NatPhy09}
\bibinfo{author}{\bibfnamefont{O.~M.} \bibnamefont{Auslaender}},
  \bibinfo{author}{\bibfnamefont{L.}~\bibnamefont{Luan}},
  \bibinfo{author}{\bibfnamefont{E.~W.~J.} \bibnamefont{Straver}},
  \bibnamefont{\textit{et~al.}},
  \bibinfo{journal}{Nature Phys.} \textbf{\bibinfo{volume}{5}},
  \bibinfo{pages}{35} (\bibinfo{year}{2009}).

\end{thebibliography}

\end{document}